\documentclass[10pt]{article}
\usepackage[cp1251]{inputenc}
\usepackage[english]{babel}
\usepackage{graphicx}
\usepackage{amssymb}
\usepackage{amsmath}
\usepackage{amsthm}
\usepackage{amsfonts}
\usepackage{amssymb}
\title{{\bf \Large  Scalar Field Cosmology in Lyra's Geometry}\\
{\normalsize ~~{\bf V.\,K. Shchigolev}\thanks{E-mail:
vkshch@yahoo.com},~~{\bf E.\,A. Semenova
\thanks{E-mail: elena.s15@mail.ru}}}\\
{\small {\it Ulyanovsk State University, 42 L. Tolstoy Str.,
Ulyanovsk 432000, Russia}}\\
\vspace{2mm}
\small \begin{quote}{\bf Abstract} --  The new classes of homogeneous cosmological models for the scalar fields are build in the context of Lyra’s geometry. The different types of exact solution for the  model are obtained by applying two procedures, viz the generating function method and the first order formalism.
 \\
\vspace{2,5mm}
{\bf PACS numbers}: 98.80.-k; 98.80.Jk; 04.20.Jb.\\
{\bf Key words}: Cosmology, Scalar Field, Phantom, Tachyon Field, Lyra's Geometry.\\
\end{quote}}
\date{}

\begin{document}

\maketitle
\vspace{-2.5cm}
\section{Introduction}

After the formulation of General Relativity (GR) by Einstein, many alternative geometric theories have been developed in order to explain gravitation phenomena. Inspired by geometrizing
gravitation, Weyl \cite{A1} proposed a more general theory in which both gravitation and electromagnetism are described geometrically.
For a long time, Weyl's theory was not taken seriously due to non-integrability of
length of vector under parallel displacement.
Later Lyra \cite{A2} suggested a modification of Riemannian
geometry by introducing a gauge function which removes the non-integrability
condition of the length of a vector under parallel transport. This modified Riemannian
geometry is known as Lyra’s geometry. In contrast
to Weyl's geometry, in Lyra's geometry the connection is metric preserving as in
Riemannian geometry, and length transfers are integrable. It should be noted that Lyra introduced a gauge function into the structure-less manifold, as a result of which a displacement field
arises naturally. Several authors (see, e.g. \cite{A3}-\cite{A12}) have studied cosmology in Lyra's geometry. This alternating theory is of interest because it produces effects similar to
those produced in Einstein’s theory

Soleng \cite{A13} has pointed out that the cosmologies based on Lyra’s manifold with constant gauge vector will either include a creation field and be equal to Hoyle’s creation field cosmology
\cite{A14,A15}, or contain a special vacuum field, which together with the gauge vector term,
may be considered as a cosmological term. Contrary to common assertion that the displacement vector field can play the role of the cosmological constant by itself, we want to emphasize that it can never play this role being alone, as seen from the equations presented below in this article.

In general relativity Einstein succeeded in geometrising gravitation by identifying
the metric tensor with the gravitational potentials. In the scalar tensor theory of
Brans-Dicke on the other hand, the scalar field remains alien to the geometry. Lyra's
geometry is more in keeping with the spirit of Einstein's principle of geometrisation
since both the scalar and tensor fields have more or less intrinsic geometrical
significance. Furthermore, the present theory predicts the same effects, within
observational limits, as far as the classical Solar System tests are concerned, as well as
tests based on the linearised form of the field equations \cite{A12}.

To date, several authors have studied cosmology in Lyra's
geometry with both a constant displacement field and a time-dependent one. For instance, in \cite{A12} the displacement field is allowed to be time dependent, and
the Friedmann-Robertson-Walker (FRW) models are derived in Lyra's manifold. Those  models are free of the big-bang singularity and solve the entropy and horizon problems which beset the standard models based on Riemannian geometry. Recently, cosmological models in the frame work of
Lyra’s geometry in different contexts are investigated in several papers (see, e.g \cite{A16}-\cite{A24}).

In last few decades there has been considerable interest in alternative theories of gravitation coursed by the investigations of inflation and, especially, late cosmological acceleration which is  well proved in many papers \cite{A25}-\cite{A30}. In order to explain so
unexpected behavior of our universe, one can modify the
gravitational theory \cite{A31}-\cite{A36}, or construct various
field models of the so-called dark energy (DE) which equation of state (EoS)
satisfies $w= p/\rho< -1/3$. Presently, there is an uprise of interest in scalar fields in
GR and alternative theories of gravitation in this context. Therefore, the
study of cosmological scalar-field models in Lyra's geometry may be relevant for
the cosmic acceleration models.

Most studies in Lyra's cosmology involve a perfect fluid.
Strangely, at least up to our knowledge, the case of
scalar field in Lyra's cosmology was not studied properly. Here we would like to fill this
gap. In this paper, we consider a scalar (quintessence or phantom) field  and a tachyon
field cosmological evolution  in the context of Lyra's geometry.
With motivation provided above, we have obtained exact solutions of
Einstein’s modified field equations for the spatially flat Friedmann metric within the frame work of Lyra’s geometry. For this purpose, we employ two methods, viz the generating function method and the first order formalism.

\section {Field equations}

The Einstein's field equations based on Lyra's manifold, as proposed in \cite{A3} and \cite{A4} in normal gauge, may be written as
\begin{equation}\label{1}
R_{ik}- \frac{1}{2} g_{ik} R + \frac{3}{2}\phi_i \phi_k - \frac{3}{4}g_{ik}\phi^j \phi_j = -\kappa^2 T_{ik},
\end{equation}
where $\phi_i$ is the displacement vector, $\kappa^2 = 8\pi G$ and other symbols have their usual meanings in the Riemannian geometry.

We assume a perfect fluid form for the energy-momentum tensor:
\begin{equation}\label{2}
T_{ik}= (\rho +p)u_i u_k -p\, g_{ik},
\end{equation}
and co-moving coordinates $u_i u^i = 1$, where $u_i = (1,0,0,0)$.
We also let $\phi_i$ be the time-like vector
\begin{equation}\label{3}
\phi_i = (\beta,0,0,0),
\end{equation}
where $\beta = \beta(t)$ is a function of time alone. The metric for FRW space-time is given by
\begin{equation}\label{4}
ds^2 = d t^2- a^2 (t)(d r^2+\xi^2 (r)d \Omega
^2),
\end{equation}
where $\xi(r)=\sin r,r,\sinh r$ \ in accordance with a sign of the
curvature $k=+1,0,-1$.
For this metric together with (\ref{2}) and (\ref{3}), the field equations (\ref{1}) become
\begin{equation}\label{5}
3H^2 + \frac{3 k}{a^2} - \frac{3}{4} \beta^2 = \kappa^2 \rho,
\end{equation}
\begin{equation}\label{6}
2 \dot H + 3H^2 + \frac{k}{a^2} + \frac{3}{4} \beta^2 = - \kappa^2 p,
\end{equation}
where $H = \dot a / a$ is the Hubble's parameter.

Equations (\ref{5}) and (\ref{6}) lead to the continuity equation as follows
\begin{equation}\label{7}
\dot \rho + \frac{3}{2 \kappa^2} \beta \dot \beta + 3 H \Big[\rho + p + \frac{3}{2 \kappa^2}\beta^2 \Big]=0.
\end{equation}
It is easy to find that the main equations of the model, i.e. (\ref{5}), (\ref{6}) and (\ref{7}), can be presented in standard GR form,
\begin{equation}\label{8}
3H^2 + \frac{3 k}{a^2} = \kappa^2 \rho_{eff},
\end{equation}
\begin{equation}\label{9}
2 \dot H + 3H^2 + \frac{k}{a^2} = - \kappa^2 p_{eff},
\end{equation}
and
\begin{equation}\label{10}
\dot \rho_{eff}+3 H \Big[\rho_{eff} + p_{eff}\Big]=0,
\end{equation}
by introducing two effective parameters:
\begin{equation}\label{11}
\rho_{eff} = \rho + \frac{3 \beta^2}{4\kappa^2}, ~~p_{eff} = p + \frac{3 \beta^2}{4\kappa^2}.
\end{equation}
For the analysis of model, it is suitable to consider
the EoS parameter $w$ and the deceleration parameter $q$ defined by
\begin{equation}\label{12}
w=\frac{p_{eff}}{\rho_{eff}}=-1-\frac{2}{3}\frac{\dot H}{H^2},\,\,\,q=- \frac{a \ddot a}{\dot a^2}=-1-\frac{\dot H}{H^2}.
\end{equation}
In the absence of matter, that is when $\rho = p =0$, the effective EoS is equal to $w = + 1$, which corresponds to the so-called stiff fluid.  That is why the displacement vector can never play the role of a cosmological term for which the EoS $w= -1$ is required.

To proceed further, we have to specify the type of scalar field. For the sake of simplicity, from now on we consider  a flat FRW cosmology: $k=0$.

\section {Quintessence (phantom) Lyra's cosmology}

In this section, we consider a quintessence (or phantom) field as a source of gravity in Lyra's cosmology. Therefore, we have for the effective parameters (\ref{11}) as follows:
\begin{equation}
\rho_{eff}=\frac{\epsilon}{2}\dot \varphi^2 + V(\varphi)+\frac{3 \beta^2}{4\kappa^2},\,p_{eff}=\frac{\epsilon}{2}\dot \varphi^2 -V(\varphi)+\frac{3 \beta^2}{4\kappa^2},\label {13}
\end{equation}
where $ \epsilon = + 1$  represents  quintessence while  $ \epsilon = - 1$ refers to phantom field. In view of (\ref{13}), the set of basic equations (\ref{8}),(\ref{9}) becomes
\begin{eqnarray}
3 H^2 &=& \epsilon\dot \varphi^2 + 2 V(\varphi)+\frac{3}{4}\beta^2, \label{14}\\
\dot H &=& - \epsilon\dot \varphi^2 -\frac{3}{4}\beta^2, \label{15}
\end{eqnarray}
where and in what follows we assume $4\pi G=1$. Even for the given potential  $V(\varphi)$ and displacement function $\beta(t)$, it is difficult to find exact solution for this model. However, a class of exact solutions can be obtained, say, in terms of the so-called generating function \cite{A37,A38}.

\subsection{The generating function method}

Since  $\beta(t)$ is an arbitrary function so far, we can transfer this arbitrariness into a new one, say coupling function $f(\varphi(t))$ as follows
\begin{equation}\label{16}
\beta^2=\frac{4}{3}f^2(\varphi)\dot \varphi^2.
\end{equation}
As a result, we have from (\ref{13}) and (\ref{16}) that
\begin{eqnarray}
\rho_{eff}= \frac{1}{2}[f^2(\varphi)+\epsilon] \dot \varphi^2 + V(\varphi),\label{17}\\ p_{eff}=\frac{1}{2} [f^2(\varphi)+\epsilon] \dot\varphi^2 -V(\varphi). \label{18}
\end{eqnarray}
Re-defying the scalar field as $
|f^2(\varphi)+\epsilon|\dot \varphi^2=\dot \psi^2,
$ i.e. $\psi=\displaystyle\pm \int \sqrt{|f^2(\varphi)+\epsilon|}\,d\varphi$,
and putting $\varepsilon = sign [f^2(\varphi)+\epsilon]$,
we get  the following expressions for the effective parameters (\ref{13}): $ \rho_{eff}= \varepsilon\frac{1}{2}\dot \psi^2 + U(\psi),~~ p_{eff}=\varepsilon\frac{1}{2}\dot\psi^2 -U(\psi)$,
where $U(\psi)=V(\varphi(\psi))$. As seen  in the effective parameters,  the sign $\varepsilon$ of kinetic term  can change (for $\epsilon=-1$) from positive to negative and vice versa.

Substituting (\ref{17}) and (\ref{18}) into Eqs. (\ref{8}), (\ref{9}), we have the following set of the main equations for our model:
\begin{eqnarray}
\label{19}  H^2&=&\frac{2}{3}\Big(\frac{1}{2}[f^2(\varphi)+\epsilon] \dot \varphi^2 + V(\varphi)\Big), \\
\label{20} \dot H&=&-[f^2(\varphi)+\epsilon] \dot \varphi^2 .
\end{eqnarray}
In order to obtain  exact solutions and following to the method of generating function \cite{A37,A38},  we assume that $F(\varphi)=\dot \varphi$. Applying the latter to the set of equations (\ref{19}),(\ref{20}), we can obtain the following general solution:
\begin{equation}\label{21}
t(\varphi)=\int \frac{d\varphi}{F(\varphi)},
\end{equation}
\begin{equation}\label{22}
H(\varphi)=-\int [f^2(\varphi)+\epsilon] F(\varphi)\,d\varphi ,
\end{equation}
\begin{equation}\label{23}
V(\varphi)=\frac{3}{2}H^2(\varphi)-\frac{1}{2}[f^2(\varphi)+\epsilon] F^2(\varphi),
\end{equation}
and
\begin{equation}\label{24}
a(\varphi)=a_0 \exp\Big(\int\frac{H(\varphi)}{F(\varphi)}d\,\varphi \Big).
\end{equation}
Moreover, as it follows from (\ref{12}) and (\ref{22}), the EoS parameter can be given by
\begin{equation}\label{25}
w=-1+\frac{2}{3}\,[f^2(\varphi)+\epsilon]\Big(\frac{F(\varphi)}{H(\varphi)}\Big)^2.
\end{equation}
Let us consider two particular cases.

\subsubsection{The case $F(\varphi)=\lambda$}
\begin{figure}[t]
\centering
\includegraphics[width=90mm,height=7cm]{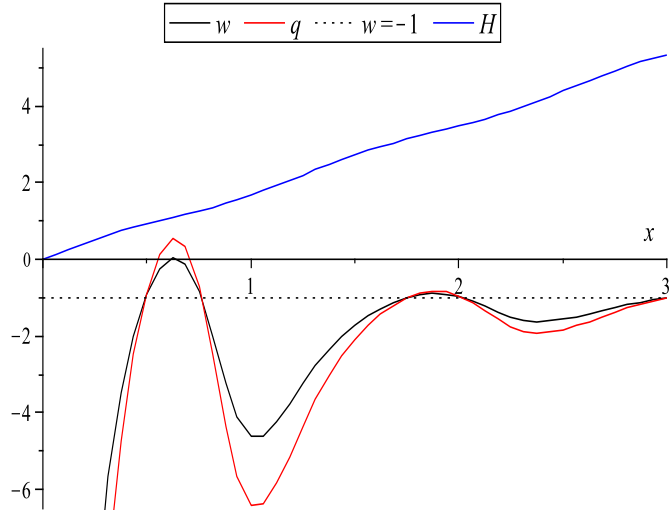}\\
\caption{The EoS parameter $w$, the deceleration parameter $q$ and the Hubble parameter $H$ in the case $F(\varphi)=\lambda$ versus $x=\lambda t$. Here, $\epsilon = -1$, $\omega_0=5$,  $f_0^2=0.56$, $\lambda=4$.}
\label{fig1}
\end{figure}
In order to obtain an explicit solution using  this method, one has to specify the coupling function $f(\varphi)$, or put some other additional condition. Here we assume that the coupling function can be represented by
\begin{equation}\label{26}
f(\varphi)=\sqrt{2} f_0 \sin\frac{\omega_0}{2}\varphi.
\end{equation}
With $f_0^2>1/2$ and $\epsilon=-1$, the kinetic term in (\ref{17}), (\ref{18}) can periodically change its signature, evolving from the phantom regime to the quintessence one and vice versa.
From Eqs. (\ref{21})-(\ref{24}), we can find that
\begin{equation}\label{27}
\varphi = \lambda t,
\end{equation}
\begin{equation}\label{28}
H(\varphi) = -\lambda \Big[(f_0^2+\epsilon)\varphi -\frac{f_0^2}{\omega_0}\sin \omega_0\varphi\Big],
\end{equation}
\begin{equation}\label{29}
a(\varphi) = a_0 \exp\Big\{-\frac{f_0^2+\epsilon}{4}\varphi^2+\frac{f_0^2}{2\omega_0^2}(1-\cos\omega_0\varphi) \Big]\Big\},
\end{equation}
and
\begin{equation}\label{30}
V(\varphi) = \frac{3 \lambda^2}{2} \Big[(f_0^2+\epsilon)\varphi -\frac{f_0^2}{\omega_0}\sin 2 \omega_0\varphi\Big]^2-\frac{\lambda^2}{2} \Big[(f_0^2+\epsilon)-f_0^2\cos \omega_0\varphi\Big].
\end{equation}
The corresponding expression for the displacement function can be found from Eqs. (\ref{16}),(\ref{26}) and (\ref{27}) as
\begin{equation}\label{31}
\beta(t)=\frac{2\sqrt{2}}{3}\lambda f_0 \sin\Big(\frac{\omega_0 \lambda t}{2}\Big),
\end{equation}
According to (\ref{12}) and (\ref{28}), we have for the EoS parameter
\begin{equation}\label{32}
w=-1 +\frac{2}{3}\frac{\epsilon+f_0^2(1-\cos\omega_0 \lambda t)}{\displaystyle \Big[(f_0^2+\epsilon)\lambda t -\frac{f_0^2}{\omega_0}\sin \omega_0 \lambda t\Big]^2},
\end{equation}
and the similar expression for the deceleration parameter $q$ with $1$ instead of the fraction $2/3$. This solution is plotted in Fig. 1. It can be seen that the model expands in acceleration. At some points,  where $\cos\omega_0\lambda t = 1-f_0^2$, the expansion is de Sitter ($q = -1$).
The model periodically crosses the phantom divide in the both direction.

\subsubsection{The case $F(\varphi)=\lambda \varphi$}

\begin{figure}[t]
\centering
\includegraphics[width=90mm,height=7cm]{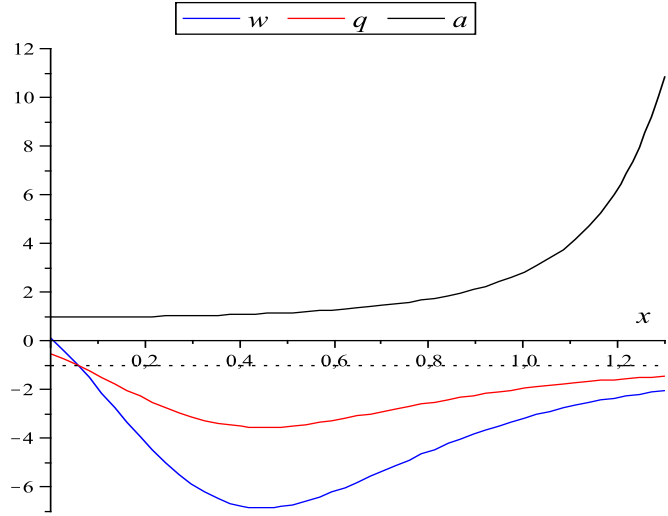}\\
\caption{The EoS parameter $w$, the deceleration parameter $q$ and the scale factor $a$ in the case $F(\varphi)=\lambda \varphi$ versus $x=\lambda t$. Here, $\epsilon = -1$, $\varphi_0 = 1$, $f_0^2=1.12$.}
\label{fig2}
\end{figure}
In order to obtain an explicit solution with some interesting feature in this case, we chose the coupling function $f(\varphi)$ in the following form:
\begin{equation}\label{33}
f(\varphi)=\frac{f_0}{\varphi}.
\end{equation}
For this choice, we have
\begin{equation}\label{34}
\varphi = \varphi_0 \exp(\lambda t)
\end{equation}
After integration, Eqs. (\ref{21})-(\ref{24}) gives
\begin{equation}\label{35}
H=-\lambda\Big[f_0^2\lambda t+f_0^2 \ln\varphi_0+\epsilon\frac{\varphi_0^2}{2}\exp(2\lambda t)\Big],
\end{equation}
\begin{equation}\label{36}
a(t)=a_0 \exp \Big\{-\frac{f_0^2 \lambda^2}{2} t^2 -f_0^2\lambda t \ln\varphi_0-\epsilon \frac{\varphi_0^2}{4}\Big(\displaystyle e^{2\lambda t}-1\Big)\Big\},
\end{equation}
where the constant of integration is chosen so that $a_0= a(0)$. Besides, we can find that the potential is given by
\begin{equation}\label{37}
V(\varphi) = \frac{\lambda^2}{2} \Big[3(f_0^2 \ln \phi+\frac{\epsilon}{2}\varphi^2)^2-\epsilon \varphi^2-f_0^2\Big].
\end{equation}
After that, the Eos parameter can be found as follows
\begin{equation}\label{38}
w=-1+\frac{2}{3}\,\frac{f_0^2+\epsilon\varphi_0^2\exp(2\lambda t)}{\Big[f_0^2\lambda t+f_0^2 \ln\varphi_0+\epsilon\frac{\varphi_0^2}{2}\exp(2\lambda t)\Big]^2}.
\end{equation}
The similar expression can be also obtained for the deceleration parameter. Some features of this solution is shown in Fig. 2.
It can be seen that this model expands in acceleration as well. At the only point $t_0=\lambda^{-1}\ln(f_0/\varphi_0)$, the expansion is de Sitter, that is $q = -1$, and the model crosses the phantom divide evolving from the quinessence sector to the phantom one.

\subsection{The superpotential method}

\begin{figure}[t]
\centering
\includegraphics[width=90mm,height=6cm]{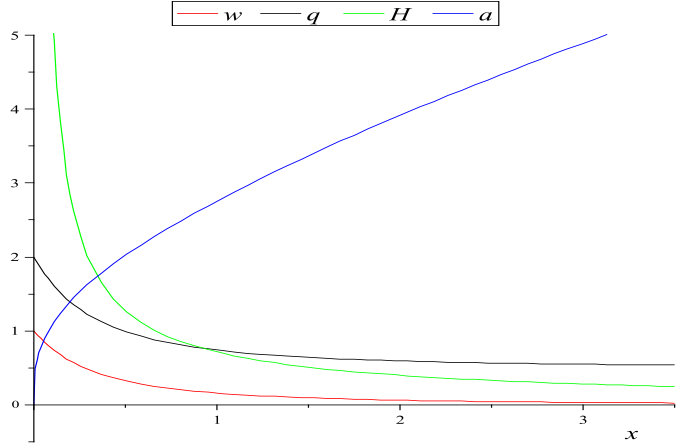}\\
\caption{The EoS parameter $w$, the deceleration parameter $q$, the scale factor $a$ and the Hubble parameter $H$ versus $x=X_0 t$ in the case $\epsilon = +1, \lambda = \sqrt{3/2}$. Here, $X_0 = 3/2$.}
\label{fig3}
\end{figure}
 Other class of exact solutions can be obtained in terms of the so-called superpotential. This procedure was first performed with a single scalar field in \cite{A39}, and it was later re-opened as "the first order formalism" and extended on the case of two or more fields in \cite{A40}. The method of  superpotential  can be effectively applied to the quintom models as well (see, e.g., \cite{A41}).
Keeping in mind the superpotential method, we now represent the geometrical field of displacement vector as a function of a new field $\alpha(t)$:
\begin{equation}\label{39}
\beta^2(t)=\frac{4}{3}\dot \alpha^2(t).
\end{equation}
Now we get the new set of equations instead of (\ref{14}),(\ref{15}):
\begin{eqnarray}
3 H^2 &=& \epsilon\dot \varphi^2 + 2 V(\varphi)+\dot \alpha^2(t), \label{40}\\
\dot H &=& - \epsilon\dot \varphi^2 -\dot \alpha^2(t), \label{41}
\end{eqnarray}
Let us introduce the superpotential function $W(\varphi, \alpha)$ by the equation
\begin{equation}\label{42}
H=W(\varphi, \alpha),
\end{equation}
in which the Hubble parameter $H(t)$, as a function of time, is presumably expressed in terms of fields $\varphi(t),\, \alpha (t)$. Substituting (\ref{42})  into (\ref{41}), one can obtain two first-order equations as follows:
\begin{equation}\label{43}
\dot \varphi =-\epsilon W_{\varphi},\,\, \dot \alpha = -W_{\alpha}.,
\end{equation}
where $W_{\varphi}=\partial W/ \partial \varphi$, $W_{\alpha}=\partial W/ \partial \alpha$.
The potential can be obtained from (\ref{40}),  (\ref{42}) and (\ref{43}) as follows
\begin{equation}\label{44}
V(\varphi)=\frac{3}{2} W^2- \frac{\epsilon}{2} W_{\varphi}^2-\frac{1}{2}W_{\alpha}^2.
\end{equation}
As the potential depends on $\varphi$ alone, we have $\partial V/\partial \alpha = 0$. Taking into account (\ref{43}) and (\ref{44}),  it is easy to show that this is equivalent to
\begin{equation}\label{45}
\frac{d W_{\alpha}}{d\,t} +3 W W_{\alpha} = 0.
\end{equation}
The latter can be satisfied in many ways. Let us now consider one simple example of exact solution.

We suppose that the superpotential is presented by
\begin{equation}\label{46}
W(\varphi, \alpha)=X(\varphi)\, Y(\alpha),
\end{equation}
that allows to rewrite (\ref{44}) in the following form:
\begin{equation}\label{47}
V(\varphi)=\frac{1}{2}\,X^2(\varphi)\Big[3 Y^2(\alpha)-\epsilon Y^2(\alpha)\Big(\frac{X'(\varphi)}{X(\varphi)}\Big)^2-Y'^2(\alpha)\Big].
\end{equation}
We consider one example of the particular solution for this model based on
\begin{equation}\label{48}
X(\varphi)=X_0 e^{-\lambda \varphi},\, \,\,Y(\alpha)=\cosh(\sqrt{\mu} \alpha),
\end{equation}
where $\mu = 3-\epsilon \lambda^2$. So we have
\begin{equation}\label{49}
\dot \varphi = -\epsilon W_{\varphi}=\epsilon\lambda  X_0  e^{-\lambda \varphi}\cosh(\sqrt{\mu}\alpha),
\end{equation}
\begin{equation}\label{50}
\dot \alpha=-W_{\alpha}=-\sqrt{\mu} X_0 e^{-\lambda \varphi}\sinh(\sqrt{\mu}\alpha),
\end{equation}
\begin{equation}\label{51}
H(t)=X_0 e^{-\lambda \varphi(t)}\cosh(\sqrt{\mu}\alpha(t)).
\end{equation}
At the same time, we have the following expression for the potential (\ref{47}):
\begin{equation}\label{52}
V(\varphi)=\frac{\mu}{2}X_0^2 \, e^{-2 \lambda \varphi}.
\end{equation}
Combining and integrating Eqs. (\ref{49}),(\ref{50}), we can get
\begin{equation}\label{53}
\exp(-\lambda \varphi) = \Big[\sinh(\sqrt{\mu}\alpha)\Big]^{\displaystyle \epsilon \frac{\lambda^2}{\mu}},
\end{equation}
where the constant of integration is chosen to be equal to zero.
Substituting the latter into  (\ref{50}), we have the following equation for $\alpha$:
\begin{equation}\label{54}
\dot \alpha = -\sqrt{\mu} X_0 \Big[\sinh(\sqrt{\mu}\alpha)\Big]^{\displaystyle \frac{3}{\mu}}.
\end{equation}
This equation can be integrated explicitly for several values of $\mu$ (or $\lambda$ and $\epsilon$).

As an example, we consider the case $\epsilon = +1, \lambda = \sqrt{3/2}$ or  $\mu = 3/2$. Integrating equation  (\ref{54}) and taking into account (\ref{39}), we can obtain that
\begin{equation}\label{55}
\alpha=\sqrt{\frac{2}{3}}\tanh^{-1}\Big[\frac{2}{3X_0 t+2}\Big],\,\, \beta^2=\frac{32 X_0^2}{[(3 X_0 t+2)^2-4]^2},
\end{equation}
where the constant of integration is chosen from  the condition $\tanh \{\sqrt{3/2}\,\alpha(0)\}= 1$.
With the help of (\ref{51}), (\ref{53}) and (\ref{55}), it is possible to find that
\begin{equation}\label{56}
H(t)=2 X_0 \frac{3 X_0 t+2}{(3 X_0t+2)^2-4},\,\,a(t)=a_0 \Big[(3 X_0 t+2)^2-4\Big]^{1/3},
\end{equation}
and
\begin{equation}\label{57}
\varphi =\frac{1}{\sqrt{6}} \ln \Big[(3 X_0t+2)^2-4\Big].
\end{equation}
From Eqs. (\ref{12}) and (\ref{56}), it can be found that
\begin{equation}\label{58}
w= \frac{4}{(3 X_0 t+2)^2},\,\,\,q=\frac{1}{2}+\frac{6}{(3 X_0 t+2)^2} .
\end{equation}
Time evolution of this model is plotted in Fig. 3.

\section{Lyra's cosmology of tachyon field}

In this section, we consider a tachyon field $\chi$ as a source of gravity in Lyra's cosmology. Substituting the well known  expressions of tachyonic $\rho$ and $p$  into (\ref{11}),  we have
\begin{equation}
\rho_{eff}=\frac{V(\chi)}{\sqrt{1-\dot \chi^2}} + \frac{3 \beta^2}{8},\,\,\,p_{eff}=-V(\chi)\sqrt{1-\dot \chi^2}+\frac{3 \beta^2}{8}.\label {59}
\end{equation}
As before, we put $H(t)=W[\chi(t),\alpha(t)]$ and suppose the substitution (\ref{39}).
Due to (\ref{59}), the set of basic equations (\ref{8}), (\ref{9})  becomes as follows
\begin{eqnarray}
\frac{3}{2} H^2 &=& \frac{V(\chi)}{\sqrt{1-\dot \chi^2}}+\frac{1}{2}\dot \alpha^2,
\label{60}\\
\dot H &=& - \frac{V(\chi)\dot \chi^2}{\sqrt{1-\dot \chi^2}} -\dot \alpha^2. \label{61}
\end{eqnarray}
By inserting $H=W(\chi,\alpha)$ into (\ref{61}), one can obtain two first-order equations,
\begin{equation}\label{62}
\dot \chi =-\frac{2W_{\chi}}{3W^2-W_{\alpha}^2},\,\, \dot \alpha = -W_{\alpha}.,
\end{equation}
where $W_{\chi}=\partial W/ \partial \chi$, $W_{\alpha}=\partial W/ \partial \alpha$.
The potential is followed from (\ref{60}) - (\ref{62}) in the form:
\begin{equation}\label{63}
V(\chi)=\frac{1}{2}\sqrt{\displaystyle \Big(3 W^2-W_{\alpha}^2\Big)^2 - 4 W_{\chi}^2}.
\end{equation}
As this potential is independent on $\alpha$, we have $\partial V/\partial \alpha = 0$. In view of (\ref{62}), (\ref{63}),  it is easy to prove that the latter is equivalent to equation (\ref{45}).

So we have a wide range of possibilities to solve the model equations assuming some certain dependence $W(\chi,\alpha)$. Instead, we can provide several classes of solution for the model evolving from some conditions on superpotential. Below, we show how it can be realized with the help of some ansatz for the superpotential.

\subsection{An example of exact solution for tachyon model}

One of the simplest ansatz for the superpotential may be written as follows:
\begin{equation}\label{64}
W_{\chi}=\lambda_1 W^m(\chi,\alpha),\,\,\,W_{\alpha}=\lambda_2 W^m(\chi,\alpha),
\end{equation}
where $m,\,\lambda_1,\,\lambda_2$ are constants, and the equality $W_{\chi,\alpha}=W_{\alpha,\chi}$ is true. Therefore in view of (\ref{45}) and (\ref{64}), we have
\begin{equation}\label{65}
H(t) = W(\chi(t),\alpha(t))=\frac{m}{3(t+t_0)},
\end{equation}
where $t_0$ is an integration constant.
Hence, equations (\ref{62}) become as follows
\begin{equation}\label{66}
\dot \chi =-\frac{2 \lambda_1 (m/3)^{m-2} (t+t_0)^m}{3(t+t_0)^{2m-2}-\lambda_2^2(m/3)^{2m-2}}\,,\,\,\dot \alpha= -\lambda_2\frac{(m/3)^{m}}{(t+t_0)^m}\,.
\end{equation}
From the last equation in (\ref{62}) and Eqs. (\ref{64}), (\ref{65}), it immediately follows that the displacement vector is given by
\begin{equation}\label{67}
\beta(t)=\lambda_2\frac{2}{\sqrt{3}}\Big(\frac{m}{3}\Big)^{m}(t+t_0)^{-m}.
\end{equation}
The EoS parameter and the deceleration parameter (\ref{12}) are constant and defined by
\begin{equation}\label{68}
w=-1+\frac{2}{m},\,\,\,q=-1+\frac{3}{m}.
\end{equation}

In order to distinguish between  various DE models, Sahni et al. \cite{A42} proposed a cosmological diagnostic pair $\{r, s\}$ called statefinder.
The statefinder test is a geometrical one based on the expansion of the scale factor a(t) near the present time $t_0$:
$$
a(t) = 1 + H_0 (t - t_0) -\frac{1}{2} q_0 H^2_0 (t - t_0)^2 +\frac{1}{6}r_0 H^3_0 (t - t_0)^3 + ...,
$$
where $a(t_0) = 1$ and $H_0, q_0, r_0$ are the present values of the Hubble parameters, deceleration parameter and the statefinder index $ r = \dddot a /a H^3$ respectively. The statefinder parameter $s$ is the combination of $r$ and $q$: $s = (r - 1)/3(q - 1/2)$.
The important feature of statefinder is that the spatially flat $\Lambda$CDM has a fixed point
$\{r, s\} = \{1, 0\}$. Departure of a DE model from this fixed point is a good way of establishing the ‘distance’ of this model from flat $\Lambda$CDM. In terms of the Hubble parameter and its derivatives with respect to cosmic time the statefinder parameters of a flat FRW model are given by
\begin{equation}\label{69}
r=1+3\frac{\dot H}{H^2}+\frac{\ddot H}{H^3},\,\,\,s= - \Big(\frac{2}{3 H}\Big)\,\frac{3 H \dot H+\ddot H}{3 H^2+2 \dot H}\,.
\end{equation}
With the help of Eqs.(\ref{65}),(\ref{69}), one can find that
\begin{equation}\label{70}
r=1-\frac{9}{m}+\frac{18}{m^2},\,\,\,s=\frac{2}{m},
\end{equation}
for the model considered. The plot of  $\{r,s\}$ points corresponding to the different values of $m$ is shown in Fig. 4.
\begin{figure}[t]
\centering
\includegraphics[width=70mm,height=5cm]{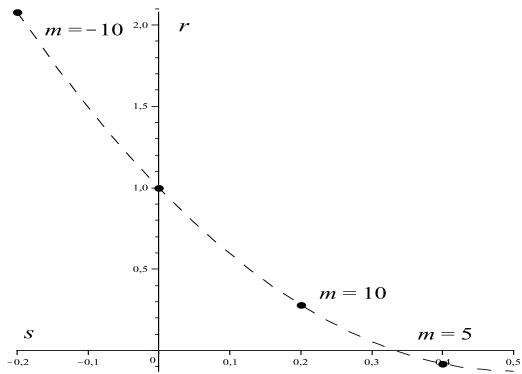}\\
\caption{The points on $\{r,s\}$ plan  for different illustrative
values of parameter $m$. The $\{1,0\}$ point is related to the location of standard
$\Lambda$CDM model.}
\label{fig3}
\end{figure}
\section{ Conclusion}

In this paper, we have studied FRW cosmological models in normal gauge for Lyra’s manifold with the quintessence (phantom) and tachyon scalar fields as the origin of gravity. We have built the new classes of FRW cosmological models of these scalar fields in the context of Lyra’s geometry. The different types of exact solution for the  model are obtained by applying two procedures: the generating function method and the first order formalism.
We hope that the derived model is the next step in the development of Lyra's cosmology, and  can be utilized to describe the dynamics of the evolution of the actual Universe.

\end{document}